\documentclass{jetpl}
\twocolumn

\lat

\title{Phase Transition in Strongly Degenerate Hydrogen Plasma}

\rtitle{Phase Transition in Dense Hydrogen\ldots}

\sodtitle{Phase Transition in Strongly Degenerate Hydrogen Plasma}

\author{V.\,S.\,Filinov$^{*}$, V.\,E.\,Fortov$^{*}$, 
M.\,Bonitz$^{**}$\thanks{e-mail: michael.bonitz@physik.uni-rostock.de}, 
and P.\,R.\,Levashov$^{*}$}

\rauthor{Filinov {\i et al.}}

\sodauthor{Filinov, Fortov, Bonitz, Levashov}

\address{$^*$Institute for High Energy Densities RAS,
127412 Moscow, Russia\\
$^{**}$Fachbereich Physik, Universit\"at Rostock, D-18051 Rostock, Germany}

\dates{20 August 2001}{*}

\abstract{Direct fermionic path-integral Monte-Carlo simulations of strongly
coupled hydrogen are presented. Our results show evidence for the hypothetical 
plasma phase transition. Its most remarkable manifestation is the appearance 
of metallic droplets which are predicted to be crucial for  
the electrical conductivity allowing to explain the rapid increase
observed in recent shock compression measurments.}


\begin{document}

\maketitle

Hydrogen at high pressures remains the subject of many investigations, see
e.g. \cite{Kraeft86, Kalman98} for an overview. Many interesting phenomena, 
such as a metal--insulator transition (MIT), Mott effect and a plasma phase
transition (PPT) have been predicted. They occur in situations where
both, quantum and Coulomb effects are important, making a theoretical 
analysis difficult. Among the most promising theoretical approaches to
such systems is the path-integral quantum Monte Carlo (PIMC) method 
\cite{Zamalin77, Ceperley96} which has seen
remarkable recent progress, e.g.  \cite{Ceperley96, Berne98}.
However, for Fermi systems, these simulations are substantially hampered by
 the so-called fermion sign problem. Additional assumptions such as fixed node
 and restricted path concepts have been introduced to overcome this difficulty
\cite{Ceperley96}. It can be shown, however, that such assumptions do not 
reproduce the correct ideal Fermi gas limit \cite{Filinov01}.

Recently, we have presented a new path integral representation which
avoids additional approximations (direct path integral Monte-Carlo, DPIMC)
which has successfully been applied to strongly coupled hydrogen 
\cite{Levashov00, FilinovPLA00, FilinovJETPL00}, see below.
In this work we apply the DPIMC method to the analysis of dense liquid 
hydrogen in the region of the hypothetic plasma phase transition 
\cite{norman-starostin,Kraeft86,schlanges,sc92}. Computing the equation of 
state and the 
internal energy, we find clear indications for the existence of the PPT --
its first confirmation by a {\em first-principle} method. It is shown that 
the PPT manifests itself by the formation of large metallic droplets which are 
crucial for the plasma transport properties.

It is well known that the thermodynamic properties of a
quantum system are fully determined by the partition function $Z$.
For a binary mixture of $N_{e}$ electrons
and $N_{i}$ protons, $Z$ can be written as

\begin{eqnarray}
\qquad Z(N_e,N_i,V,\beta) &=&
{Q(N_e,N_i,\beta)}/{N_e!N_i!}
\nonumber\\
\qquad Q(N_e,N_i,\beta) &=&
\sum_{\sigma}\int\limits_V dq \,dr
\,\rho(q, r,\sigma;\beta).
\label{q-def}
\end{eqnarray}
Here,  $q\equiv \{{\bf q}_1, {\bf q}_2, \ldots, {\bf q}_{N_i}\}$ are
the coordinates of the protons and and
$\sigma = \{\sigma_1, \ldots, \sigma_{N_e}\}$ and
$r \equiv \{{\bf r}_1, \ldots, {\bf r}_{N_e}\}$
are the electron spins and coordinates,
respectively, and $\beta=1/k_{B}T$.
The density matrix $\rho$ in Eq.~(\ref{q-def}) is
represented in the common way by a path integral \cite{Feinman65}:

\begin{eqnarray}
&&\rho(q,r,\sigma;\beta) =
\frac{1}{\,\lambda_i^{3N_i}\lambda_{\Delta}^{3N_e}} \sum_{P} (\pm 1)^{\kappa_P}
\int\limits_{V} dr^{(1)} \dots dr^{(n)}\times
\label{rho-pimc}
\\\nonumber
&& \rho\left(q,r,r^{(1)};\Delta\beta\right) \, \dots \,
\rho\left(q,r^{(n)},{\hat P} r^{(n+1)};\Delta\beta\right)
\,{\cal S}(\sigma, {\hat P} \sigma'),
\end{eqnarray}
where $\Delta\beta\equiv\beta/(n+1)$ and
$\lambda^{2}_{\Delta}=2\pi\hbar^2\Delta\beta/m_e$.
Further, $r^{n+1}\equiv r^{n}$, $\sigma'=\sigma$; i. e., the electrons
are represented by fermionic loops with the coordinates
(beads) $[r]\equiv [r, r^{(1)},\dots, r^{(n)}, r]$.
The electron spin gives rise to the spin part of the density matrix ${\cal S}$,
whereas exchange effects are accounted for by the permutation operator
${\hat P}$
and the sum over the permutations with parity $\kappa_P$.

Following \cite{Zamalin77}, we use a modified representation
(\ref{rho-def})
of the high-temperature density matrices on the r.h.s. of
Eq.~(\ref{rho-pimc}) which is suitable for efficient direct
fermionic PIMC simulation of plasmas. With the error of the
order $\epsilon \sim (\beta Ry)^2 \chi/(n+1)$ vanishing with a
growing number of beads, we obtain the approximation
\begin{eqnarray}
&&\sum_{\sigma}\rho(q,r,\sigma;\beta) =
\frac{1}{\,\lambda_i^{3N_i}\lambda_{\Delta}^{3N_e}}
\sum\limits_{s=0}^{N_e} \rho_s(q,[r],\beta),
\label{rho-def}
\\
&&\rho_s(q,[r],\beta) = \frac{C^s_{N_e}}{2^{N_e}}\,
e^{-\beta U(q,[r],\beta)} \prod\limits_{l=1}^n
\prod\limits_{p=1}^{N_e} \phi^l_{pp}
{\rm det} \,|\psi^{n,1}_{ab}|_s,
\nonumber\\
&&U(q,[r],\beta) = U^i(q) +
\sum\limits_{l=0}^n
\frac{U^e_l([r],\beta)+U^{ei}_l(q,[r],\beta)}{n+1},
\nonumber
\end{eqnarray}
where $\chi$ is the degeneracy parameter and $U^i$, $U^e_l$ and $U^{ei}_l$
denote the sum of the binary Kelbg potentials $\Phi^{ab}$
\cite{Kelbg,ppcf01} between protons, electrons at vertex ``$l$'', and
electrons (vertex ``$l$'') and protons, respectively.

In Eq.~(\ref{rho-def}),\, $\phi^l_{pp}\equiv \exp[-\pi |\xi^{(l)}_p|^2]$
arises from the kinetic energy part of the density matrix of the electron
with index $p$, and we introduced dimensionless distances between
neighboring vertices on the loop, $\xi^{(1)}, \dots \xi^{(n)}$.
Finally, the exchange matrix is given by
$$
||\psi^{n,1}_{ab}||_s\equiv \left|\left|
  \exp\left\{-\frac{\pi}{\lambda_{\Delta}^2}
  \left|(r_a-r_b)+ y_a^n\right|^2\right\}
\right|\right|_s,
$$
$$
\mbox{with} \quad  y_a^n=\lambda_{\Delta}\sum_{k=1}^{n}\xi^{(k)}_a,
$$
where the subscript $s$ denotes the number of electrons having the same
spin projection.
From the above expressions (\ref{q-def})--(\ref{rho-def})
one readily computes the internal energy and the equation of state:
\begin{eqnarray}
 \beta E &=& \frac{3}{2}(N_e+N_i)-\beta\frac{\partial\ln Q}{\partial\beta},
\label{e}
\\
 \beta p &=& \frac{\partial\ln Q}{\partial V} =
 \left[\frac{\alpha}{3V} \frac{\partial\ln Q}{\partial\alpha}\right]_{\alpha=1}.
\label{p}
\end{eqnarray}

In our simulations we used $N_e=N_i=50$ and $n=20$. To test the MC procedure,
we have considered a mixture of {\em ideal} degenerate electrons and 
classical protons for which the thermodynamic quantities are known analytically.
The agreement, up to the degeneracy parameter $\chi$ as large as 10,
has been very good and improved with increasing number of
particles \cite{Levashov00}. Further, the method was successfully 
tested in applications to electrons in a harmonic trap 
\cite{afilinov-etal.prl01}.
For the case of {\em interacting} electrons
and protons in dense hydrogen we have previously performed a series of 
calculations over a wide range of the classical coupling parameter
 $\Gamma$ and degeneracy
$\chi$ for temperatures $T\ge10,000$~K. The analysis of the results has 
clearly shown a number of interesting phenomena, such as
formation and decay of hydrogen bound states
 \cite{FilinovPLA00, FilinovJETPL00, ppcf01}, 
 including hydrogen atoms, molecules, molecular ions, clusters and futher, 
at high densities, pairing of electrons
and ordering of protons into a Wigner crystal
\cite{FilinovJETPL00}.

In this work we present new results which concentrate on the 
hypothetical plasma phase transition \cite{norman-starostin}. 
For this purpose, we analyze the plasma properties and compute 
the equation of state (\ref{p}) and internal energy (\ref{e}) of dense 
hydrogen along two isotherms, $T=10^{4}$~K and $5\cdot 10^{4}$~K.
Fig.~1 shows pressure and energy vs. density at
$T=5\cdot 10^{4}$~K. For comparison, we also include the results 
for an ideal plasma. As expected, due to Coulomb interaction and 
bound state formation, the nonideal plasma results are 
below the ideal ones. We mention that our results are in a good 
agreement with
restricted path integral calculations of Militzer and 
Ceperley (Fig.1a contains available data points for a slightly higher temperature
of $6.25\cdot 10^4$~K \cite{Militzer01}. For higher temperatures the 
agreement is very good \cite{ppcf01}. Most importantly, at this temperature 
the pressure increases monotonically with density, and at high densities
a continuous increase of the degree of ionization (Mott effect) is 
found.

However, at $T=10^{4}$~K the properties of the hydrogen plasma
change qualitatively, cf. Figs.~2,3. While the overall trend of the 
pressure, Fig.~2.a, is still a monotonic increase, in the density region
of $0.1 \dots 1.5$~g/cc the plasma exhibits unusual behavior. Inside 
this region the Monte Carlo simulations do not converge to an 
equilibrium state, the pressure strongly fluctuated 
and reached even negative values.
Such a behavior is typical for Monte-Carlo simulations of metastable
systems. Note that no such peculiarities appear for densities below 
and above this interval as well as for the isotherm 
$T=5\cdot 10^{4}$~K and for higher temperatures. 

These facts suggest that our simulations have encountered the 
plasma phase transition predicted by many chemical models of 
partially ionized hydrogen, e.g. 
\cite{norman-starostin,Kraeft86,schlanges,sc92}.
According to these models, this is a first-order transition with 
two coexisting phases of different degree of ionization. While 
canonical Monte Carlo simulations do not yield the coexisting phases and 
the coexistence pressure directly, they allow to analyze in detail the 
actual microscopic particle configurations. A typical particle arrangement 
inside the instability region, $T=10^{4}$~K and $\rho=0.3346$~g/cc, is 
shown in Fig.~3. Obviously, the protons arrange themselves into large 
clusters (droplets) with the electrons (the piecewise linear lines 
show their closed fermionic path) being fairly delocalized over the 
cluster. This is a clear precurser of the metal-like state which is 
found in the simulations for densities above the instability region.

As mentioned above, the plasma phase transition appears in many
chemical models in the same density-temperature range. However, these 
simple approaches become questionable in the region of pressure 
ionization and dissociation where a consistent treatment of all 
possible pair interactions, including charge-charge, neutral-neutral and 
charge-neutral, is crucial. Furthermore, these approaches neglect
larger bound aggregates such as clusters which our simulations reveal 
to be crucial in the metastable region. We mention that indirect 
indications for a phase transition have been found in recent density
functional studies \cite{Xu98}. In this work 
the thermodynamic properties of hydrogen in the metallic phase 
were computed (see data points in Fig.~2) and  
enhanced long-wavelength ion density fluctuations were observed 
as the density was reduced to $\rho=0.799$~g/cc 
(the lowest density explored). This 
led to unusual behavior of the ion-ion structure factor
and the effective potential which the authors of Ref.~\cite{Xu98} 
interpreted as a possible precursor 
to an incipient metal-to-insulator transition. 

Our simulations suggest that the existence of the PPT should have
a noticable influence on the transport properties. In fact, when 
the density changes from  $0.1 \dots 1.5$~g/cc, hydrogen transforms  
from a neutral into a metallic fluid. Accordingly, the electrical 
conductivity should increase rapidly. Indeed, shock compression experiments 
have revealed a dramatic increase of the
electrical conductivity by 4--5 orders of magnitude  in  a very narrow
density range of 0.3--0.5~g/cc, \cite{Nellis96, Ternovoi99}. Theoretical 
models so far cannot reproduce
this behavior correctly, predicting either a too early (hopping 
conductivity in the molecular fluid) or too late (free electron 
conductivity) increase \cite{redmer01}. But as the experimental data
(black circles and crosses in Fig.~2.a) are located right inside
the PPT region, one has to take into account a third 
conductivity mechanism -- charge transport via electron hopping between 
individual metal-lik droplets. Obviously, this mechanism will be effective 
inbetween the regions where the two other effects dominate and thus 
should allow for a much better agreement with the experiments.

Finally, we mention that our simulations predict a PPT for 
pure hydrogen plasma only. In contrast, no PPT was found for 
a binary mixture of 25\% of helium and 75\% of hydrogen (cf. Fig.~2). 

In summary, we presented direct path integral Monte Carlo 
simulations of dense fluid hydrogen in the region of the 
metal--insulator transition (MIT). Our results show evidence for the 
plasma phase transition which, to the best of our knowledge, is 
its first prediction by a {\em first principle} theory.
Most importantly, we found clear evidence for the formation of 
large metallic droplets which are predicted to play a crucial role in
transport and optics in the region of the MIT at low temperatures.
In further investigations we will focus
on a more precise analysis of the MIT 
and the plasma phase transition, including determination of its critical 
point and of the transport and optical properties of the droplets.

We are greatful to W.~Ebeling, D.~Kremp, and W.\,D.~Kraeft and S.\,Trigger 
for stimulating discussions. 
We also wish to thank D.~Ceperley and B.~Militzer
 for useful critical remarks.

Fig.~1. (a) Pressure and (b) internal energy for hydrogen 
plasma at $T=5\cdot 10^{4}$~K vs. density. 1~--- direct
PIMC simulation of this work, 2~---ideal plasma, 3~--- restricted PIMC
computations at $T=6.25\cdot 10^{4}$~K \cite{Militzer01}.
\\

Fig.~2. (a) Pressure (1-5) and electrical conductivity (6,7)
and (b) internal energy for hydrogen 
 at $T=10^{4}$~K vs. density. 1~--- direct
PIMC simulation of this work, 2~---ideal plasma, 3~--- direct PIMC
simulation of a mixture consisting of 25\% helium and 75\% hydrogen, 4~---
density functional theory \cite{Xu98}, 5~--- restricted PIMC
computations  \cite{Militzer01}, 6, 7~--- electrical conductivity of 
hydrogen (right axis), 6~--- \cite{Nellis96}, 7~--- \cite{Ternovoi99}.
\\

Fig.~3. Snapshot of a Monte-Carlo cell at $T=10^{4}$~K and
$\rho=0.3346$~g/cc. Black circles are protons, dark and light broken
lines are representations of electrons as fermionic loops with 
different spin projections.

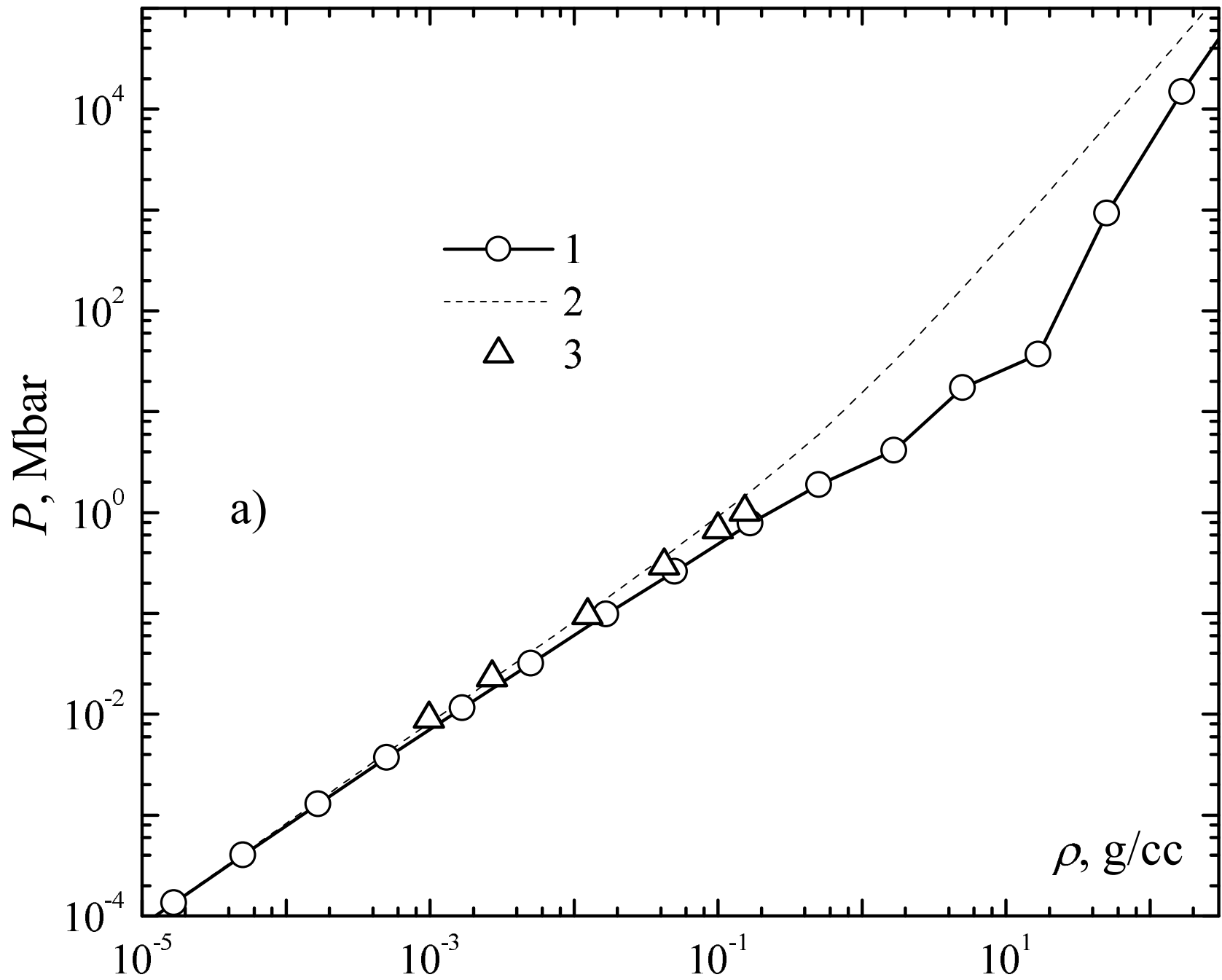
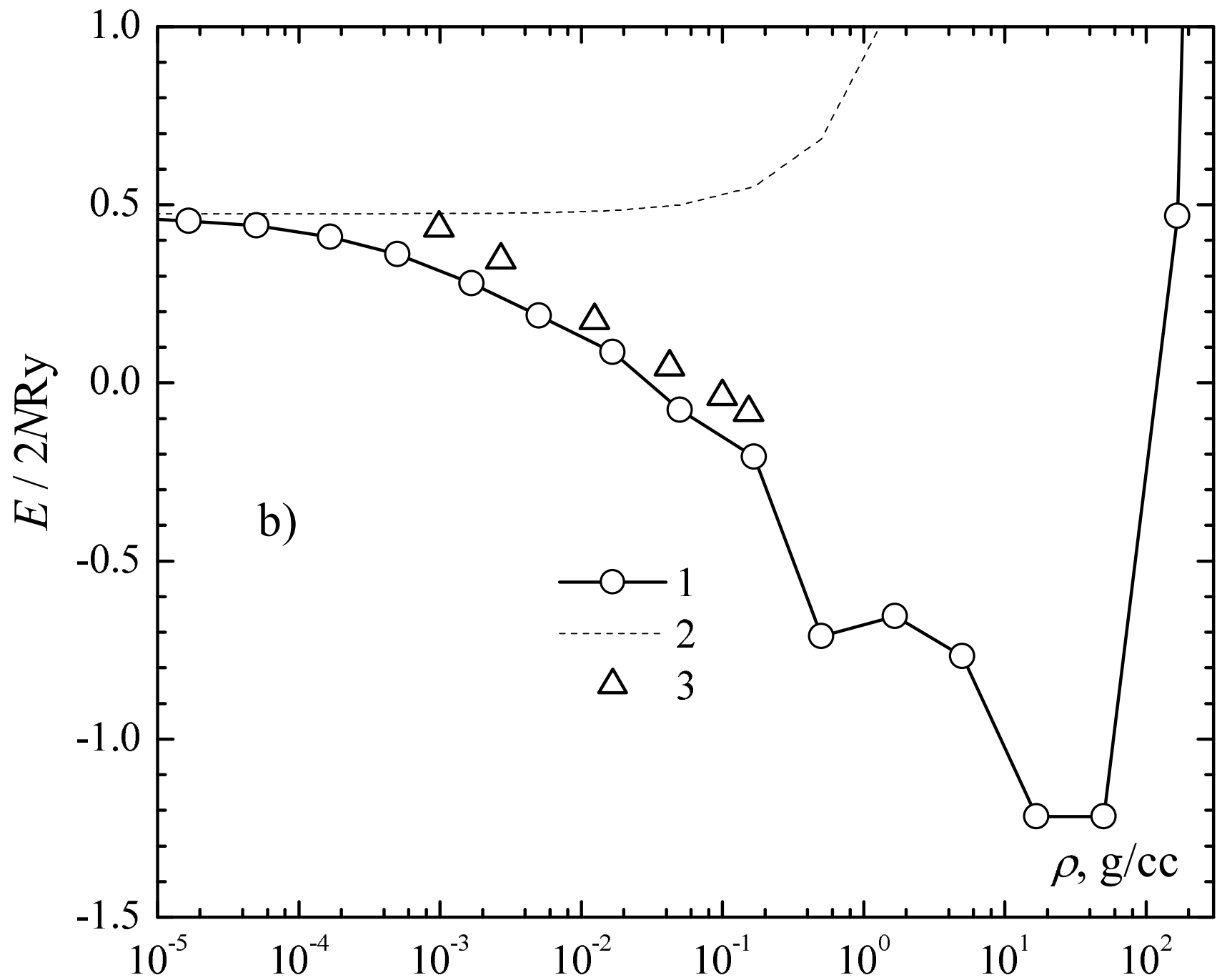
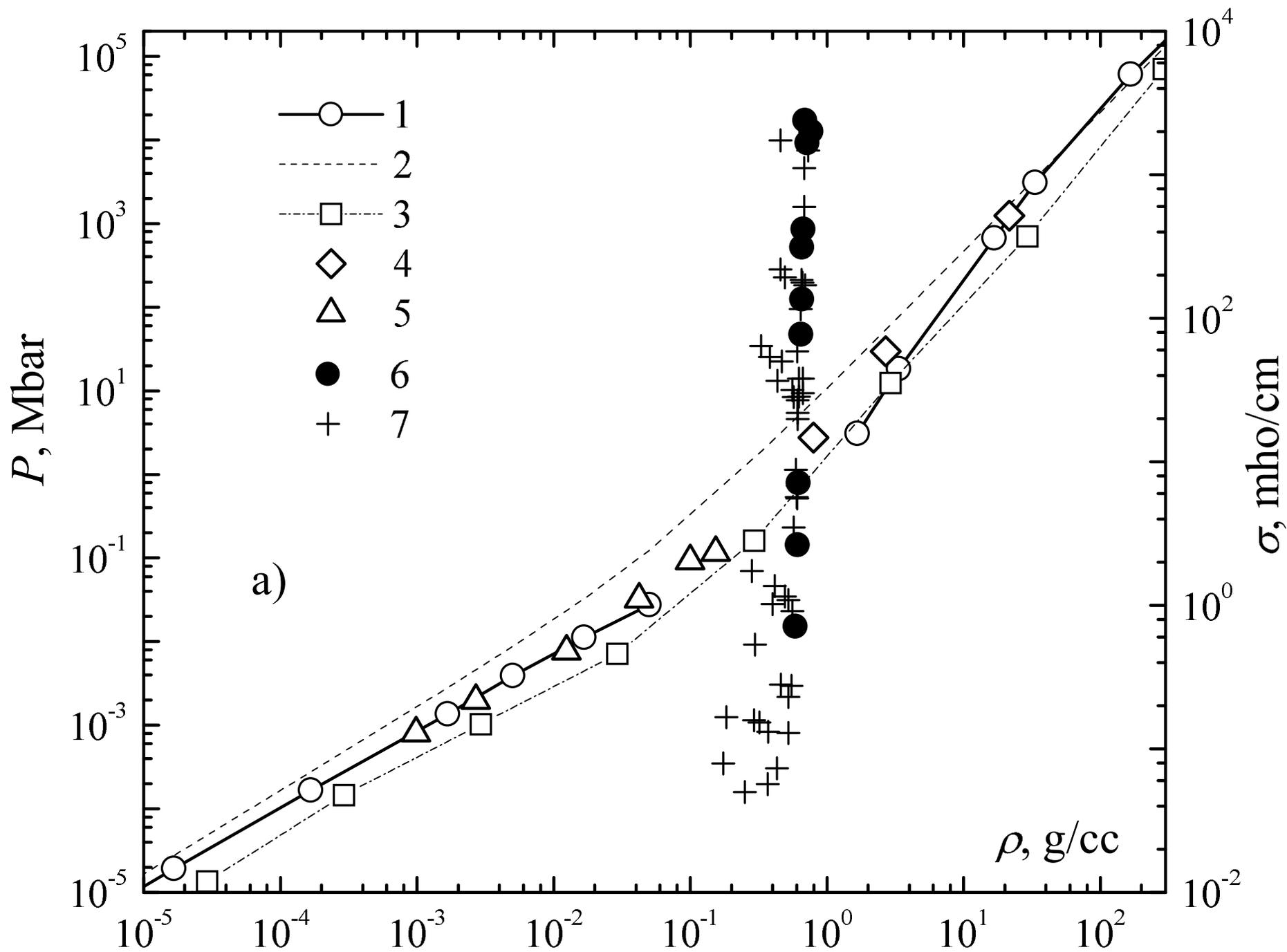
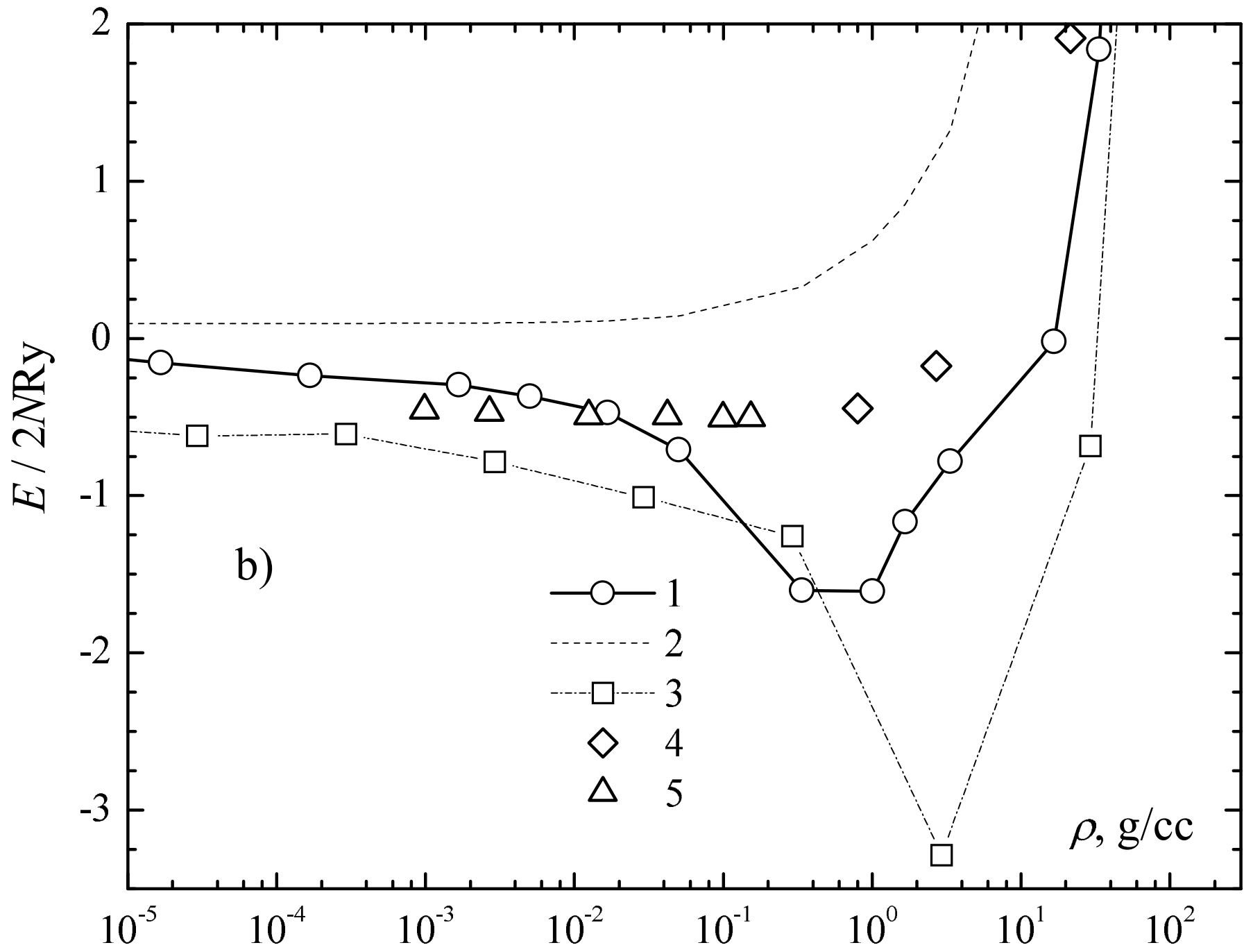
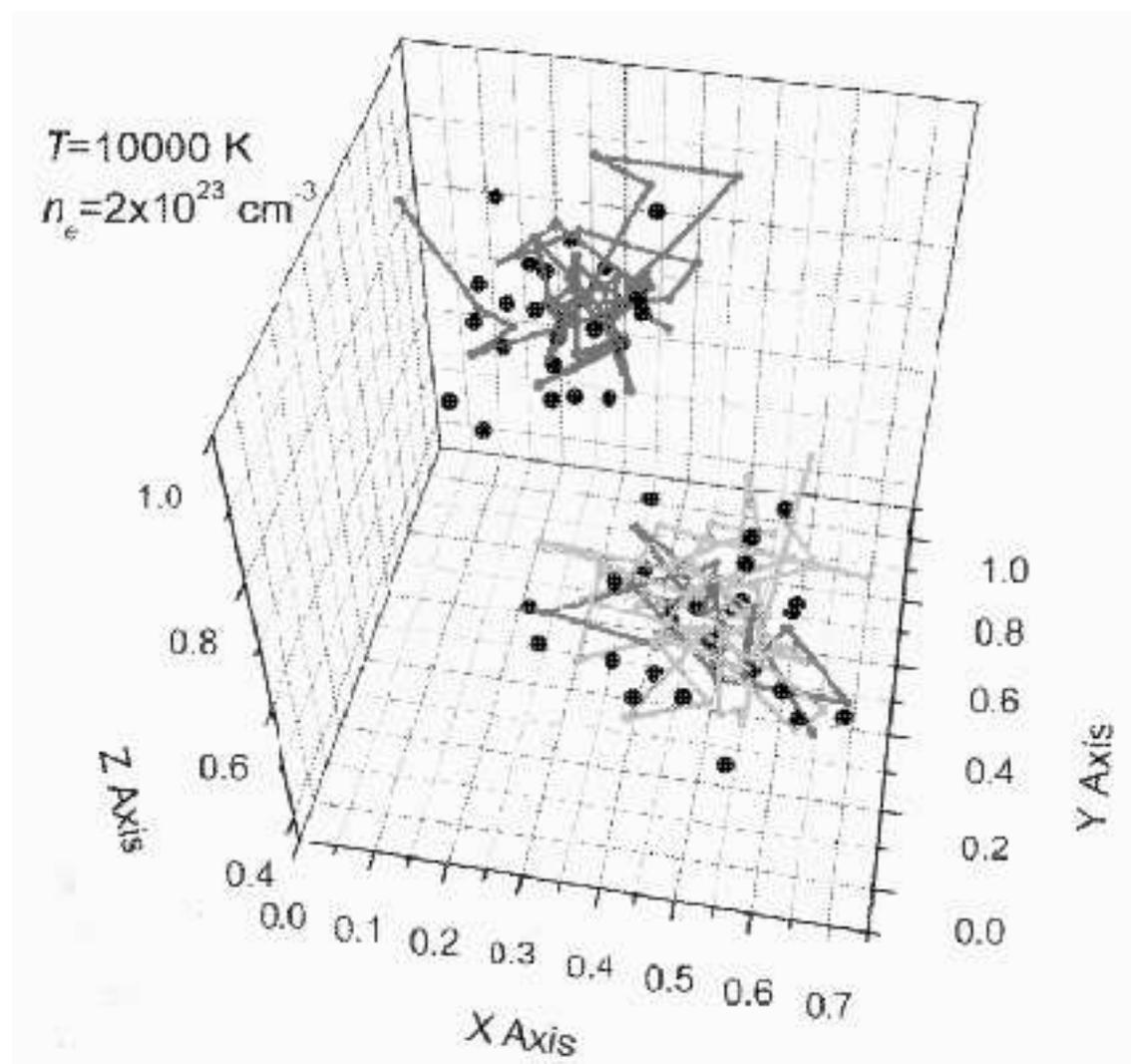
\end{document}